\documentclass[reprint,aip,cha,groupedaddress,showpacs,amsmath,amssymb]{revtex4-1} 

\usepackage{color}
\usepackage{graphicx}
\newcommand{\un}[1]{\ensuremath{\, \mathrm{#1}}}

\begin{document}

\title{Ultra-high-frequency piecewise-linear chaos using delayed feedback loops}
\author{Seth D. Cohen, Damien Rontani, and Daniel J. Gauthier}
\affiliation{Department of Physics, Duke University, Durham, North Carolina 27708, USA}
\date{\today}

\begin{abstract}
We report on an ultra-high-frequency ($ > 1$ GHz), piecewise-linear chaotic system designed from low-cost, commercially available electronic components. The system is composed of two electronic time-delayed feedback loops: A primary analog loop with a variable gain that produces multi-mode oscillations centered around 2 GHz and a secondary loop that switches the variable gain between two different values by means of a digital-like signal. We demonstrate experimentally and numerically that such an approach allows for the simultaneous generation of analog and digital chaos, where the digital chaos can be used to partition the system's attractor, forming the foundation for a symbolic dynamics with potential applications in noise-resilient communications and radar.  
\end{abstract}

\maketitle
\begin{quotation}
Recently, Corron \textit{et al.} realized a piecewise-linear electronic circuit that simultaneously generates an analog chaotic waveform and an associated digital waveform that represents the system's symbolic dynamics. Their technique is data-efficient for chaos-based applications in that the symbolic dynamics is sufficient for specifying the underlying deterministic features of the chaotic carrier signal, but their implementation was limited to the audio frequency range (20 Hz - 20 kHz). Here, we propose a new circuit architecture that operates at ultra-high-frequencies (0.3 - 3 GHz). Our approach is based on two time-delayed electronic feedback loops, one analog and the other digital. Together, they ensure an increase in complexity of the analog signal and enable fast control over its dynamics, so that the overall system still behaves in a piecewise-linear fashion. Our architecture represents a potential breakthrough in the design of inexpensive and robust electric circuits that generate piecewise-linear chaos. 
\end{quotation}

\section{Introduction}
A piecewise-linear (PWL) operator with two-states is defined as
\begin{equation}
f(x) = \left\{
     \begin{array}{lr}
       f_{1}(x), \text{ if } x < I \\
       f_{2}(x), \text{ if } x \geq I
     \end{array}
   \right.,
\label{eq:general_PWF}
\end{equation}
where $f_{1}(x)$ and $f_{2}(x)$ are linear operators with respect to $x$ and $I$ is a threshold. Everytime the value of $x$ crosses $I$, $f(x)$ switches between $f_1(x)$ and $f_2(x)$. This definition extends naturally to multiple states and thresholds.

Piecewise-linear operators have been used with discrete maps,\cite{Yeou1999} continuous systems,\cite{Konishi2011, Thus2010} and to describe certain classes of electronic circuits \cite{Vandewalle2003} and regulatory systems.\cite{Oktem2005} They introduce a nonlinear component in the dynamics of systems (called PWL systems), a minimum requirement for potentially leading to chaos. For example, when a PWL operator controls the parameters of discrete maps and linear differential equations, chaos can exist in the temporal evolution of the systems' state variables.\cite{Schulman1983} Recent studies also demonstrate that a two-state PWL operator can generate low-dimensional chaos if the system has hysteresis\cite{Saito1995} or uses multiple thresholds.\cite{Senthilkumar2005,Han2007} Such theoretical evolutions, refered to as \textit{PWL chaos}, can also be observed in experimental systems. 

Creating an experimental embodiement of a PWL system resides in the ability to design multiple linear systems (or a single system with variable parameters) and a physical switch. Examples of linear systems used to produce experimental PWL dynamics are LRC (inductance, resistance, capacitance) audio-frequency circuits.\cite{Buscarino2011} They are popular because their parameters can be easily switched with conditions based on their voltage or current. This allows the LRC circuits to operate in two different linear states.\cite{Srinivasan2011} Physical switches implemented in PWL systems use various components such as digital logic gates and operational amplifiers.\cite{Corron2010} Though the switching times between the linear states are not infinitely fast, they are usually considered negligible for LRC circuits (being orders of magnitude faster than the system's dynamics) and thus make LRC systems good for PWL experiments.\cite{Saito1981, Saito1995} 

A novel LRC-based implementation of a PWL system has been designed recently by Corron \textit{et al.}.\cite{Corron2010} In their circuit, negative damping (gain greater than one) induces oscillations about a piecewise-constant voltage with a growing amplitude that is bounded by a piecewise constant driving voltage. The driving voltage switches its sign in response to a guard condition on the system's present dynamics, and leads to the generation of audio-frequency PWL chaos. Such chaos is fully characterized by the temporal evolution of the piecewise-constant voltage that serves as an easily-accessible, real-time symbolic dynamics. This approach, with its simultaneous generation of analog and digital chaos, demonstrates advantages over traditional chaos generators from an application point-of-view. One of them is the existence of a matched filter for robust chaos-based communications\cite{Corron2011} and radars\cite{Blakely2010} in noisy environments.

The range of applications for PWL chaos, as envisioned by Corron \textit{et al.}, motivates an ultra-high-speed version of their circuit. However, non-ideal behaviors of LRC circuits make it difficult to realize an ultra-high-speed PWL mode-of-operation. Above a certain frequency, the propagation times of signals through LRC circuits are no longer negligible,\cite{Yehea2000} and the dynamics evolve on a time scale that is comparable to the switching times of the system's electronic PWL operators. Furthermore, nonlinearities tends to arise in LRC systems at ultra-high-frequency (UHF) \cite{Tamasevicius2006, Jiang2009} and thus introduce undesired effects into the PWL system. Therefore, in order to overcome these obstacles, a new architecture must addresses these challenges that are at the forefront of nonlinear dynamics and electronic design.

\begin{figure}[t!]
\begin{center}
 \resizebox{7cm}{!}{\includegraphics{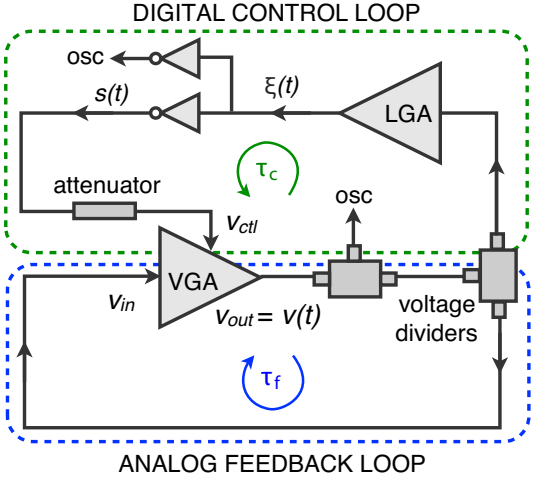}}
\end{center}
\caption{\label{fig:Experimental_Setup_2} Piecewise-linear chaotic waveform generator. An analog feedback loop connects the output voltage ($v_{\text{out}})$ of a variable-gain amplifier (VGA, Hittite HMC287MS8) and its input voltage ($v_{\text{in}}$). The VGA output also feeds a digital control loop with a logarithmic amplifier (LGA, Analog Devices AD8319) and a transistor-transistor-logic (TTL, Texas Instruments SN74AUC1G04) inverter gate in series (a second TTL gate is driven in parallel in order to monitor $s(t)$ with correct impedance matching). We monitor the signals $v(t)$ and $s(t)$ using an oscilloscope (osc).}
\end{figure}

In this paper, we present the design and operating characteristics of a novel chaotic system, shown in Fig. \ref{fig:Experimental_Setup_2}, that uses feedback loops to realize UHF low-dimensional chaos. In the following sections, we first describe the two main components of our electronic system: (i) an analog time-delayed feedback-loop, which generates a multi-mode oscillations; and (ii) a digital time-delayed control loop, which bounds the amplitude of the oscillations using a threshold condition and generates a digital-like waveform that follows the threshold crossings. Then, we show that the dynamics produced are chaotic and PWL using a time-delayed return map. Finally, we present a PWL physical model that captures the dynamics displayed by our chaos generator. 

\section{Analog Multi-mode Feedback Loop}
The time-delayed feedback loop in See Fig. \ref{fig:Experimental_Setup_2} is used to generate growing-amplitude oscillations, similar to the negative damping in Corron \textit{et al.}'s designs.\cite{Corron2010, Corron2012} It comprises a variable gain amplifier (VGA) that is self fedback. The VGA produces a voltage $v_{\text{out}} = v(t)$ that is delayed by time $\tau_{\text{f}}$ by means of coaxial cables and connected to its input [($v_{\text{in}}=v(t-\tau_{\text{f}})$)]. The net gain $g(t)$ of the feedback loop depends on the the VGA's gain, which is controlled by the voltage $v_{\text{ctl}}$ applied to its control port. If $v_{\text{ctl}}>0.5\un{V}$ then $g(t)<1$, otherwise $g(t)>1$. As we will show, these two gain values yield different dynamical regimes for the entire feedback loop. 

The presence of a time-delayed feedback loop has important implications for the spectral properties of our system compared to that of Corron \textit{et al.}. For example, it generates multi-mode oscillations instead of mono-mode oscillations.\cite{Corron2010,Corron2011} Within the bandwidth $B$ of the VGA, resonances that lead to multi-mode oscillations occur approximately at integer multiples of $1/\tau_{\text{f}}$. If the pass-band of $B$ is larger than the (approximate) inter-mode spacing, the system becomes multi-mode.

\begin{figure}[b!]
\begin{center}
\resizebox{8.5cm}{!}{\includegraphics{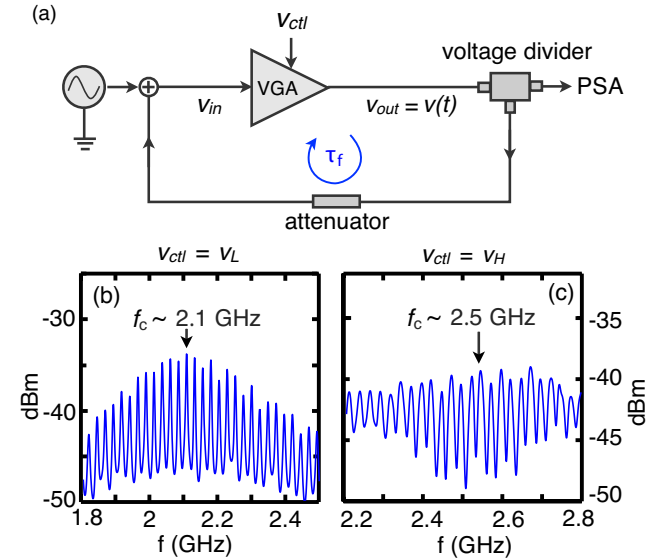}}
\end{center}
\caption{\label{fig:Experimental_Setup} Multimode characteristics of the feedback loop. (a) Experimental setup to determine the open-loop transfer function of the feedback loop. The input of a VGA is fed by a sinusoidal waveform generator and a delayed version of its output $v(t-\tau_{\text{f}})$, where the signal $v(t)$ is monitored using a power spectrum analyzer (PSA). For this measurement, an attenuator of gain $g_{\text{a}}$ in the feedback loop to maintain the net-gain $g_{\text{a}} g(t)$ below unity, avoiding self-oscillation and saturation at the resonant frequencies. In our experiment, we observe multimode characteristics with (b) $v_{\text{ctl}}=v_{\text{L}}$, the high-gain state, and (c) $v_{\text{ctl}}=v_{\text{H}}$, the low-gain state, where both states allow for the existence of up to ${\sim}20$ modes (mode spacing ${\sim}1/\tau_{\text{f}} \sim 25\un{MHz}$).}
\end{figure}

In addition, the multi-mode resonances and $B$ are dependent on the value of $v_{\text{ctl}}$. To characterize these spectral changes, we measure the feedback-loop transfer function for two different values of the voltage control port $v_{\text{ctl}}$, where the setup is shown in Fig. ~\ref{fig:Experimental_Setup}a. It comprises a sinusoidal waveform generator with automatic frequency sweep, which drives the feedback loop, and a power spectrum analyzer (PSA) that records the  output power. The experiment is first realized with $v_{\text{ctl}}=v_{\text{L}}<0.5\un{V}$, shown in Fig. \ref{fig:Experimental_Setup}b. It confirms the existence of multiple modes within $B \sim 0.4\un{GHz}$ centered at $f_{\text{c}} \sim 2.1\un{GHz}$. For $v_{\text{ctl}}=v_{\text{H}}>0.5\un{V}$ (Fig. \ref{fig:Experimental_Setup}c), we observe resonances with relatively constant amplitude, and we approximate $B \sim 0.6\un{GHz}$ with a shift in the central frequency $f_{\text{c}} \sim 2.5\un{GHz}$ (See Appendix).

\begin{figure}[t!]
\begin{center}
\resizebox{8.5cm}{!}{\includegraphics{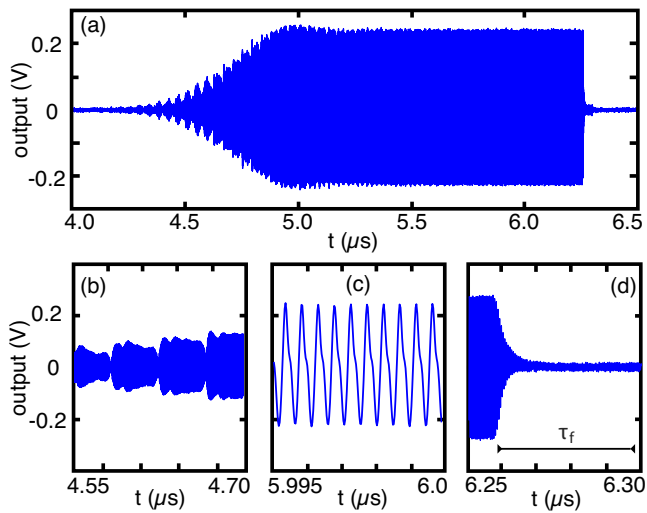}}
\end{center}
\caption{\label{fig:Growth_and_Decay} Growth, saturation, and decay of $v(t)$. (a) The feedback loop is initialized with $g_{\text{H}} = 1.5$ at $t \sim 4\un{\mu s}$. Electrical noise is amplified and band-pass filtered such that multimode oscillations emerge. Saturation occurs at $t \sim 4.85\un{\mu s}$ and then the gain is switched low ($g_{\text{L}} = 0.2$) at $t \sim 6.25\un{\mu s}$, where oscillations decay in steps of $\tau_{\text{f}}$ (given by a horizontal bar in (d)). Zooms of $v(t)$ showing regions where its amplitude is (b) governed by the system's multi-mode transfer function, (c) saturates, and (d) decays. Note that we choose $g_{\text{H}}$ and $g_{\text{L}}$ such that the rate of growth is slower than the rate of decay ($g_{\text{H}}-1 < 1-g_{\text{L}}$).}
\end{figure}

When the net-gain of the feedback is greater than unity (without an external source), the VGA amplifies multi-mode signals after each roundtrip in the feedback loop until it saturates. Figure \ref{fig:Growth_and_Decay} illustrates how the feedback loop amplifies and band-pass filters electrical noise after propagating several time through the VGA with loop net-gain $g(t)=g_{\text{H}} > 1$. The output voltage $v(t)$ starts showing modulated oscillations (see Fig. \ref{fig:Growth_and_Decay}b). When oscillations grows beyond ${\sim}0.3\un{V}$, the VGA saturates and behaves nonlinearly such that the dynamics become a stable periodic oscillation of constant amplitude at frequency ${\sim}f_{\text{c}}$ (see Fig. \ref{fig:Growth_and_Decay}c). We maintain the saturated behavior for approximately $1\un{\mu s}$ (${\sim}25\times \tau_{\text{f}}$) to confirm it is not transient dynamics. Then, the net-gain net gain is switched to $g(t)=g_{\text{L}} < 1$, and the oscillations in the feedback loop decay until they reach the noise floor of the VGA.

This experiment shows that a single feedback-loop system can generate multi-mode UHF oscillations. Away from saturation, the feedback loop operates linearly and is governed solely by the values of its net gain ($g_{\text{L,H}}$), the time delay $\tau_{\text{f}}$, and the band-pass characteristics appropriate for a given value of $v_{\text{ctl}}$. This experiment also shows that, without proper control over the time spent in the growth regime, the system saturates rapidly ($<1\mu$s), which introduces nonlinearity into the dynamics.

However, if the VGA gain is switched fast enough and with proper timing, it can avoid saturation and remain in its linear mode-of-operation. For this, in the upcoming section, we introduce an additional time-delayed feedback loop that bounds the amplitude of the multi-mode oscillations. This digital \textit{control loop} is specifically designed to parallel aspects of the guard condition from Ref. [\onlinecite{Corron2010}], where it uses a switching condition\cite{Lu2002,Liu2006} that produces a digital-like waveform $s(t)$. 

\section{Digital Control Loop}

We design our control scheme to vary $v_{\text{ctl}}$ in response to a signal that monitors the amplitude of the multi-mode oscillations in $v(t)$. When the amplitude or envelope of $v(t)$ grows beyond (decays below) a set threshold, the control loop changes the value of $v_{\text{ctl}}$ and switches the system to its decay (growth) regime. In our design, we use a logarithmic amplifier (LGA) to detect the envelope of $v(t)$ and a transitstor-transistor logic (TTL) inverter gate as a digital switch (See Appendix for characterization of these components). As a whole, the control-loop generates a digital signal and is designed to prevent saturation in the feedback loop.

As shown in Fig. \ref{fig:Experimental_Setup_2}, the feedback and control loops are combined by means of a voltage divider, so that $v(t)$ drives the VGA, the LGA, and the TTL gate. The output of the TTL gate generates a digital signal $s(t)$ that is fed into a fixed voltage attenuator connected directly to $v_{\text{ctl}}$. We use a fixed voltage attenuator to map the low and high state from $s(t)$ ($s(t)=0\un{V}$, and $2\un{V}$, respectively) onto high and low values desired for $g(t)$ ($v_{\text{ctl}}= v_{\text{L}}$ and $v_{\text{H}}$, respectively). This design is adequate to realize a switching condition for PWL dynamics, except that it requires additional time-delay to guarantee proper control on the fast dynamics existing in the feedback loop.  

With low-speed systems, the processing time of the components in a control loop, called the control-latency-time (CLT), is usually orders of magnitude faster than the timescales of the dynamics to be controlled and therefore can be neglected.\cite{Saito1995,Tsubone1998,Corron2010} In our UHF system, however, this is not the case. The average period in $v(t)$ is $0.5\un{ns}$, which should be compared to the summed CLT of the LGA and TTL gate of approximately $9\un{ns}$. As a result, it is not possible for the switching condition to affect instantaneously the feedback-loop oscillations as they are generated. 

However, we can control the envelope of the oscillations (evolving on a slower time-scale) if $s(t)$ is applied to the VGA at the proper time with respect to feedback-loop delay time. This is obtained by adjusting the time delays of both the control and feedback loops to be approximately equal. Such an approach has been already used successfully to stabilize a very-high-frequency optoelectronic chaotic system.\cite{Blakely2004}

The time delay $\tau_{\text{f}}$ of the feedback loop is tunable with respect to the length of coaxial cable used in the design of our circuit. First, we guarantee that $\tau_{\text{f}}$ is larger than the CLT and choose $\tau_{\text{f}} \sim 41\un{ns}$. Then, we use coaxial cables of time delay $\tau_{\text{coax}}$ in the control loop to approximately match the propagation time of the control signal with $\tau_{\text{f}}$. We denote this net control-loop delay time by $\tau_{\text{c}} = \text{CLT} + \tau_{\text{coax}}$ (see Fig. \ref{fig:Experimental_Setup_2}) and adjust its value to be $\tau_\text{c} \sim 40\un{ns}$ (See Appendix for a detailed description of the total control loop $\text{CLT}$).

In the following sections, we analyze the dynamics in $v(t)$ from the full system (shown in Fig. \ref{fig:Experimental_Setup_2}) and its relation to the digital output of the control loop $s(t)$.  

\section{Full System Results}

We initialize the full system with $v_{\text{ctl}} = v_{\text{L}}$ and $g(t) = g_{\text{H}}$ to ensure oscillation growth. We monitor the temporal evolution of $v(t)$ and $s(t)$ using an 8-GHz-analog-bandwidth 40-GS/s oscilloscope (DSO80804A) and analyze separately the time series and spectral content of both signals. Along with our analyses, we explain the observed dynamics based on the system design.

\subsection{Experimental time series}

\begin{figure*}[t!]
\begin{center}
 \resizebox{16cm}{!}{\includegraphics{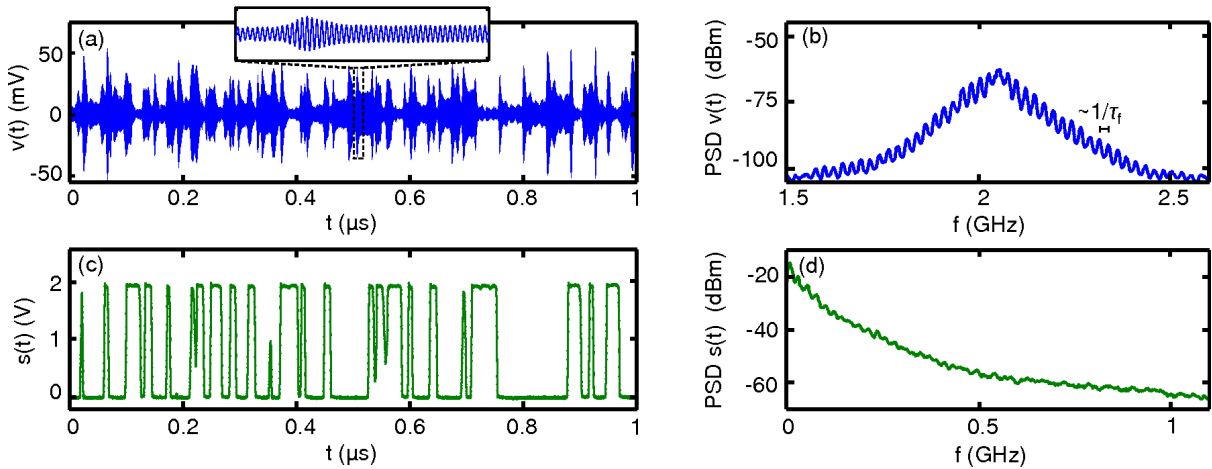}}
\end{center}
\caption{\label{fig:VS_Time_Series} Experimental time series and power spectral densities (PSD's). (a) Time series of $v(t)$ showing UHF oscillations with complex amplitude modulations. A zoom of $v(t)$ is shown for $t = 0.5-0.525$ $\mu$s. (b) PSD of $v(t)$. (c) Time series and (d) PSD of the switching state $s(t)$. The PSD's in (b) and (d) have a resolution bandwidth of 80 kHz and are smoothed over a window of 10 MHz. We note that the dynamics for this system are stable for extended periods of time, even when no thermal stabilization is used to prevent parameter drifts.}
\end{figure*}

After intial transients, the system dynamics show oscillations in $v(t)$ with a time-varying amplitude and digital-like switchings in $s(t)$. As shown in Fig. \ref{fig:VS_Time_Series}a, the evolution of $v(t)$ is aperiodic and its amplitude remains approximately between $\pm 50$ mV, within the VGA's linear mode-of-operation. The amplitude modulation in $v(t)$ leads to spectral broadening of its power spectral density (PSD) with a peak at ${\sim}f_{\text{c}}$ and multi-mode power symmetrically distributed, as shown in Fig. \ref{fig:VS_Time_Series}b. The small peaks in the PSD are spaced by ${\sim}1/\tau_{\text{f}}$. From the peak at $f_{\text{c}}$, the PSD shows a 10-dB-bandwidth of ${\sim}100$ MHz (consistent with the bandwidth of the LGA). Hence, the dynamics of $v(t)$ have a UHF broadband spectrum.

The complexity of the dynamics in $v(t)$ has two different origins: (i) multi-mode feedback and (ii) long memory.  First, noise seeds the transient oscillations of the system, which are amplified and filtered based on the system's multi-mode characteristics (see the Appendix for the initial, transient dynamics). Then, these transient amplitude modulations are stored in the system's memory as they propagate around the feedback loop and continuously the future dynamics at time $t + \tau_\text{f}$, causing the dynamics to reside in an infinite-dimensional phase space and allowing for the existence of chaos.\cite{Farmer1982}

Simultaneously, the control loop generates $s(t)$, a digital-like, asynchronous (unclocked) voltage. It switches aperiodically between its low and high states with minimum pulse widths of approximately 10 ns (see Fig. \ref{fig:VS_Time_Series}c). The finite bandwidth of $s(t)$ is visible on the experimental PSD spanning frequencies from dc to ${\sim}100$ MHz, the cuttoff frequency at -10 dB (see Fig. \ref{fig:VS_Time_Series}d). The signal $s(t)$ also shows short-pulse rejection,\cite{Zhang2009} a mechanism by which the TTL gate generates intermediate voltages in response to fast fluctuations at its input. As a result, our system shows slight deviations from ideal PWL behavior, discussed in the following section.

We finish this section by comparing the time series of $v(t)$ and $s(t)$. Unlike the audio-frequency PWL system presented in Ref. [\onlinecite{Corron2010}], there is no obvious correlation between the analog and digital dynamics. This feature is unique to our system and is caused by the multi-mode feedback loop. In the next subsection, we construct a time-delayed return map of the analog dynamics to better understand the connection between $v(t)$ and $s(t)$ and reveal the symbolic nature of $s(t)$. The return map also provides a method for calculating a Lyapunov exponent to confirm the presence of chaos. 

\subsection{Experimental Return map}

\begin{figure}[t!]
\begin{center}
 \resizebox{7cm}{!}{\includegraphics{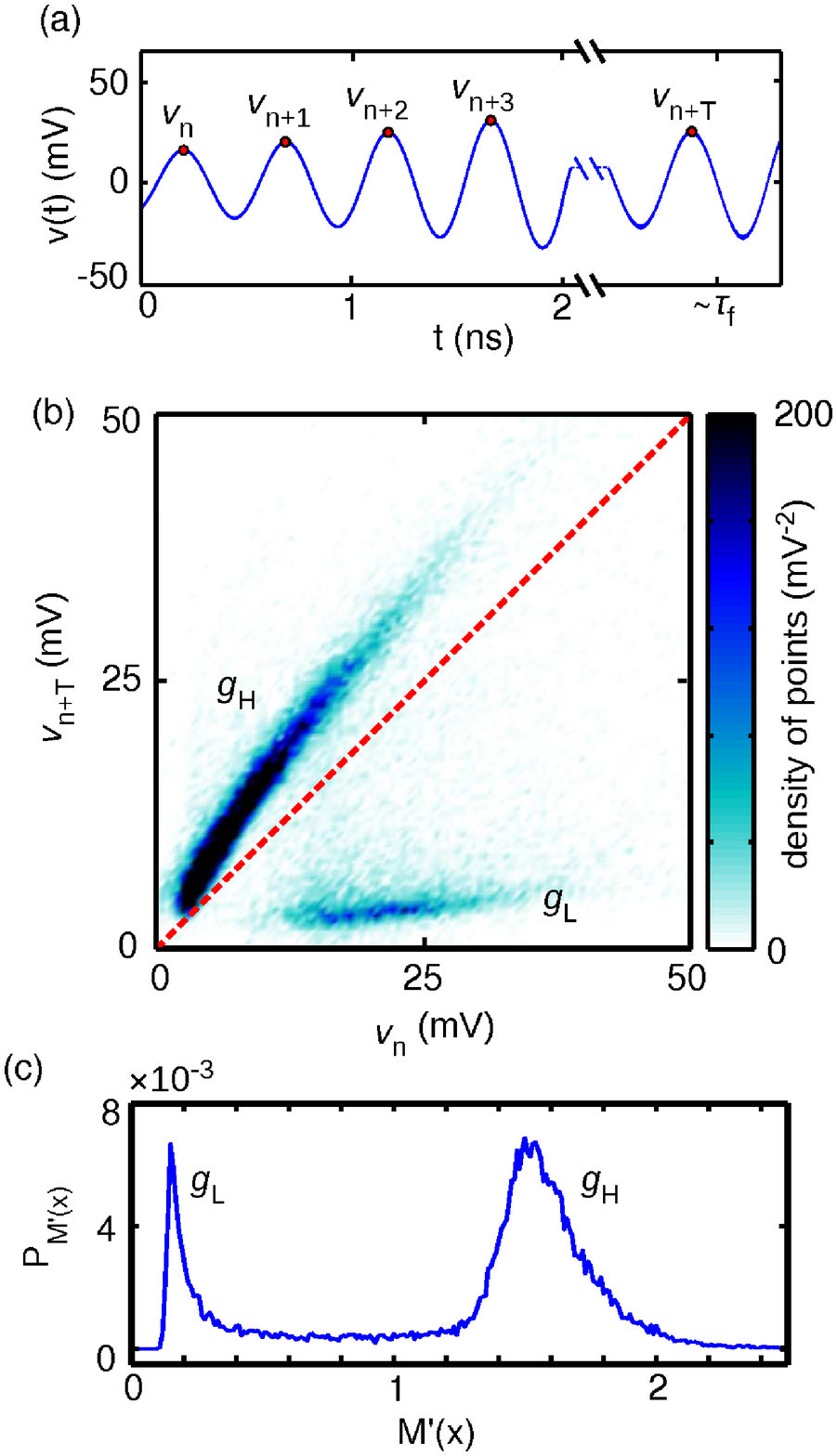}}
\end{center}
\caption{\label{fig:Exp_Maps} Construction of the experimental return map. (a) A small segment indicating the local maxima $v_{n}$ (circles) to $v_{n+T}$, where $T = 87$. (b) Return map of $v_n$. The dashed line indicates a reference line with slope equal to 1. (c) The probability density $P_{M'(x)}$ as a function of $M'(x)$. The two primary peaks in $P_{M'(x)}$ occur at $M'(x) \sim g_{\text{L,H}}$, the gain-states of the system.} 
\end{figure}

In our system, shown in Fig. \ref{fig:Experimental_Setup_2}, we note that
\begin{equation}
v(t) = g(t)v(t - \tau_{\text{f}}).
\label{eq:delay_in_out}
\end{equation}
We construct a return map by discretizing Eq. (\ref{eq:delay_in_out}). Similar to Ref. [\onlinecite{Corron2010}], we use the local maxima in $v(t)$, labeled as $v_n$, where $n$ is an integer incremented over each maxima, that are spaced by ${\sim}1/f_{\text{c}}$, as shown in Fig. \ref{fig:Exp_Maps}a. Also shown is the local maxima $v_{n+T}$, where $T$ is the integer number of local maxima that are contained in the feedback loop. Using these points, Eq. (\ref{eq:delay_in_out}) becomes a delayed return map, defined as
\begin{equation}
v_{n+T} = M(v_{n}),
\label{eq:Map}
\end{equation}
where $M(x) = g_n x$ and $g_n$ is the net gain of the system at time $t_n$.

The map is respresented graphically by a density plot in the plane of ($v_{n},v_{n+T}$). Figure \ref{fig:Exp_Maps}b shows an attractor of two different clusters of points with finite widths and slopes that coincide approximately with the gains $g_{\text{H}}$ and $g_{\text{L}}$ (labeled accordingly in the figure). The linearity of these clusters is expected because the derivative of the map $M'(x) = dM(x)/dx = g_n$. The finite widths relate to small over and under shoots in the values of the gain because of imperfections in the VGA gain and intermediate $s(t)$ values applied to $v_{\text{ctl}}$. In addition, depending on the state of $s(t)$, the map's dynamics reside on one of its two clusters, respectively. Everytime $s(t)$ switches, the map's dynamical state ($v_{n},v_{n+T}$) transitions from one cluster to the other, demonstrating that $s(t)$ partitions the map's attractor (See the Appendix for a preliminary analysis of the alphabet in the symbolic dynamics and its underlying grammar).

We also analyze the map by looking at the statistics of its derivative. In Fig. \ref{fig:Exp_Maps}c, the probability density of $M'(x)$, labeled by $P_{M'(x)}$, is given as a function of $M'(x)$ and shows two primary peaks that occur at approximately $g_{\text{L,H}}$, the most probable values of $g_n$. There is also a higher probability to have $M'(x) >$ 1 because of the larger area under the $g_{\text{H}}$ peak, which indicates on average more growth than decay in the dynamics.

Based on the finite switching times in $s(t)$, we expect finite, but small, probabilities $P_{M'(x)}\neq g_{\text{L,H}}$, which is verified in the plot. However, the slow drop-off in $P_{M'(x)}$ for $M'(x) > g_{\text{H}}$ shows that there is a significant probability to have a gain $g_n > g_{\text{H}}$. This shows that the VGA has a slightly asymmetric response for switching $v_{\text{ctl}}$ between $v_{\text{L}}$ and $v_{\text{H}}$. Together, these features reveal the effects of the slight deviations from an exactly PWL system.

To conlude our analysis, we use the probability density of $M'(x)$ to calculate the Lyapunov exponent of the map (similar to Ref. [\onlinecite{Ott2002}]) given by
\begin{equation}
h = \int \! \text{ln}|M'(x)|P_{M'(x)}\, \mathrm{d} x.
\label{eq:Lyapunov}
\end{equation}
We note that, for $\text{ln}|M'(x)| > 0$, $M'(x) > 1$ corresponds to exponential growth in the map's dynamics. Using Eq. (\ref{eq:Lyapunov}), we calculate $h = (7.2 \pm 0.2) \times 10^{-3}$, where the uncertainties are due to experimental noise and $h > 0$ is an indication of chaos (using surrogate data of equal length, we have verified that $h$ is significantly positive). In this case, $h$ describes the average rate of divergence between any given pair ($v_n,v_{n+T}$). For the divergence rate between $v_n$ and a neighboring point $v_{n+1}$, we scale $h \rightarrow h/T$ with $T=87$ such that $h/T = (8.3 \pm 0.2) \times 10^{-5}$. With both Lyapunov exponents positive, this confirms the presence of chaos in our experimental system. 

\section{Piecewise-linear Physical Model}

In this section, we model both the feedback and control loops using using delay differential equations (DDE's). All of the model parameters are obtained from regression analysis using experimental data as detailed in the Appendix.

\subsection{Multi-mode Feedback Loop Model}

We model the feedback loop as a band-pass filter with time-delayed feedback and gain \cite{Udaltsov2002}
\begin{equation}
\dot{v}(t) +\Delta(t)v(t)+\omega^{(o)}(t)\int \! v(t') \, \mathrm{d} t'=g(t)v(t-\tau_{\text{f}}),
\label{eq:BP_feedback}
\end{equation}
where $\tau_{\text{f}}$ is the feedback-loop time delay and $\Delta(t)$, $\omega^{(o)}(t)$, and $g(t)$ are \textit{approximately piecewise-constant} (APC) parameters that continuously switch. We note that Eq. (\ref{eq:BP_feedback}) is completely linear for fixed $\Delta(t)$, $\omega^{(o)}(t)$, and $g(t)$, as we have defined the feedback loop without a nonlinear saturation term. Therefore, for $g(t) > 1$, Eq. (\ref{eq:BP_feedback}) becomes unstable and $v(t)$ diverges. As we will discuss in subsection B, the control loop, which continuously switches the values of the APC parameters, is the only mechanism that keeps $v(t)$ from diverging (experimental saturation). 

The first two APC parameters are modeled as $\omega^{(o)}(t) = 2\pi\sqrt{f^{(+)}(t)f^{(-)}(t)}$ and $\Delta(t) = 2\pi(f^{(+)}(t)-f^{(-)}(t))$, where $f^{(+)}$ and $f^{(-)}$ are the upper ($+$) and lower ($-$) cutoff frequencies of the multi-mode bandpass filter (-3 dBm drop off), respectively. In the model, these cuttoff frequencies shift depending on the switching state $s(t)$ as
\begin{equation}
f^{(+,-)}(t) = \frac{s(t)}{A}\left(f^{(+,-)}_{\text{L}}-f^{(+,-)}_{\text{H}}\right)+f^{(+,-)}_{\text{H}},
\label{eq:piecewise_parameters_1}
\end{equation}
where $f^{(+)}_{\text{H}} = 2.41 \pm 0.01 \un{GHz}$, $f^{(-)}_{\text{H}} = 1.85 \pm 0.01 \un{GHz}$, $f^{(+)}_{\text{L}} = 3.29 \pm 0.1 \un{GHz}$, and $f^{(-)}_{\text{L}} = 1.97 \pm 0.04 \un{GHz}$ and $s(t)/A$ is goverened by the model for the control loop. 

The last APC parameter is the gain $g(t)$, which is also dependent on $s(t)/A$ through the relation
\begin{equation}
g(t) = \frac{s(t)}{A}\left(g_{\text{L}}-g_{\text{H}}\right)+g_{\text{H}},
\label{eq:piecewise_parameters_2}
\end{equation}
where $g_{\text{H}} = 1.5$ ($g_{\text{L}} = 0.2$) is the high (low) gain state. In our model, all of the APC constants have the same rise and fall times (instead of infinitely fast switches) that follow the physical switching state $s(t)$. This approximation captures some of the non-ideal behaviors of the experiment while maintaining approximate PWL dynamics.  

\subsection{Control Loop Model}

The control loop is modeled using two separate nonlinear operators to approximate the LGA and TTL gate. The LGA first rectifies a delayed version of $v(t)/4$, provided by the feedback loop, with a logarithmic response in amplitude and saturation for high inputs using the nonlinear function
\begin{equation}
F(v(t-\tau_{\text{c}})) = C_1 - C_2 \left(\text{tanh}\left[\text{log}_{10}\left|\frac{C_3}{4}v(t-\tau_{\text{c}})\right|\right]+1\right),
\label{eq:F}
\end{equation}
with parameters $C_1 = 1.68 \pm 0.01$ V, $C_2 = 3.24 \pm 0.09$ V, $C_3 = 24.1 \pm 5.6$ V$^{-1}$, and $\tau_{\text{c}} = 39.9 \pm 1$ ns. To finish, the output of this function drives a first order low-pass filter
\begin{equation}
\dot{\xi}(t) = -2 \pi f_{\text{L}}\xi(t)+2 \pi F(v(t-\tau_{\text{c}})),
\label{eq:Log_Amp_Eq}
\end{equation}
where $f_{\text{L}} = 0.02 \pm 0.01$ GHz is the low-pass cut-off frequency. The output $\xi(t)$ is smoothed to recover the envelope of $v(t)$. This completes the model of the LGA. 

The LGA output $\xi(t)$ then drives the TTL gate, which we approximate as a continuous nonlinear switching function
\begin{equation}
s(t) = \frac{A}{2}(1+\tanh[m(I-\xi(t))]),
\label{eq:S}
\end{equation}
where $A = 2.00 \pm 0.04$ V, $I = 0.96 \pm 0.01$ V, and $m = 51.00 \pm 0.01$ V$^{-1}$. Equation (\ref{eq:S}) switches with rise and fall times proportional to $m$ and asymptotically approaches $s(t) = 0$ and $s(t) = A$ for inputs $\xi(t)$ above and below the threshold $I$, respectively (Recall that the value of $s(t)/A$ is also used in Eqs. (\ref{eq:piecewise_parameters_1}) - (\ref{eq:piecewise_parameters_2})). 

\subsection{Simulation Results}

\begin{figure*}[t!]
\begin{center}
 \resizebox{16cm}{!}{\includegraphics{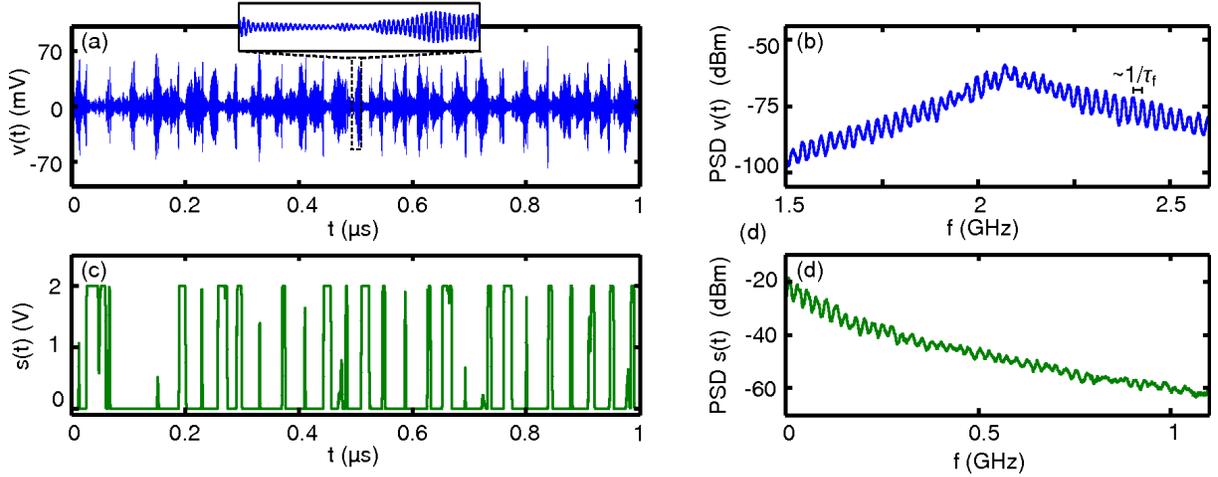}}
\end{center}
\caption{\label{fig:VS_Time_Series_SIM} Dynamics from the model. (a) Time series of $v(t)$ with a zoom shown for $t = 0.5 - 0.525$ $\mu$s and (b) PSD of $v(t)$. (c) Time series and (d) PSD of $s(t)$. The integration was performed using initial conditions for $t \leq 0: s(t) = 0, \xi(t) = 1.5$ and $v(t)$ = Gaussian white noise with a variance $\sigma = 10^{-9}$ $\text{V}^2$ and sampled at a fixed time-step $\delta t = 5$ ps. No additional noise is introduced at later times into the model. The PSD's are plotted assuming 50 $\Omega$ impedance with a frequency resolution of 80 kHz and averaged over a window of 10 MHz.}
\end{figure*}

\subsubsection{Numerical time series}

We integrate Eqs. (\ref{eq:BP_feedback})-(\ref{eq:piecewise_parameters_2}) using a third-order Adams-Bashforth algorithm. The simulated time series and frequency spectra for $v(t)$ and $s(t)$ are plotted in Fig. \ref{fig:VS_Time_Series_SIM}, which mirrors the format of Fig. \ref{fig:VS_Time_Series} for easy comparison with the experimental results.

As shown in Figs. \ref{fig:VS_Time_Series_SIM}a-b, the central frequency of $v(t)$ is ${\sim}f_{\text{c}}$, and the amplitude of $v(t)$ is bounded between $\pm 80$ mV. Similar to the experiment, the amplitude modulations of $v(t)$ are aperiodic and the PSD of $v(t)$ shows a multi-mode power spectrum. The dynamics of the model are thus a good approximation to the experiment, producing a qualitatively similar, broadband, UHF, multi-mode $v(t)$. 

Recall that the model does not include any saturation, and thus, in order for such dynamics in $v(t)$ to remain bounded, the control loop must produce $s(t)$ to swtich the gain. In Fig. \ref{fig:VS_Time_Series_SIM}c, $s(t)$ is an asynchronous, digital-like waveform with some instances of short-pulse rejection due to the continuous nature of Eq. (\ref{eq:S}). Different from the experiment, the PSD of $s(t)$ shows small oscillations in its amplitude. We hypothesize that these oscillations are due to high-frequencies in $v(t)$ that bleed through the model's filter for the LGA. A higher order low-pass filter in the model can suppress these oscillations but requires an increase in the model's complexity.

Next, we examine the PWL nature of the physical model by constructing a time-delayed return map of the integrated dynamics in $v(t)$.

\subsubsection{Simulated return map}
\begin{figure}[h!]
\begin{center}
 \resizebox{7cm}{!}{\includegraphics{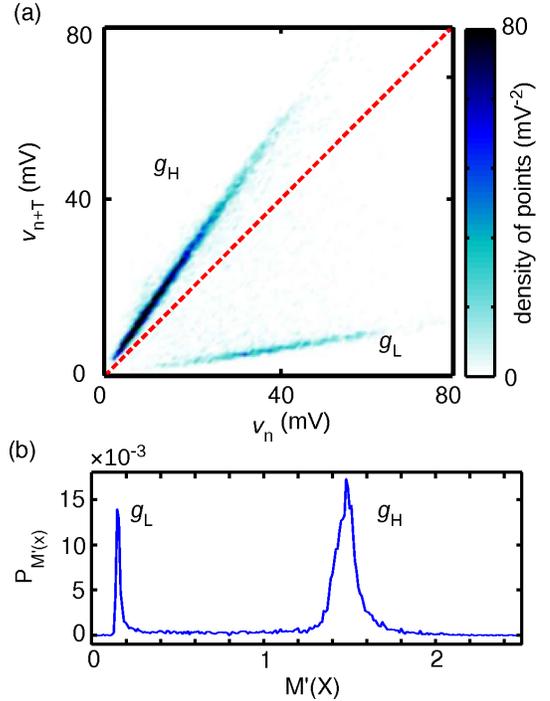}}
\end{center}
\caption{\label{fig:Exp_Maps_SIM} (a) Return map from simulation constructed from $v_n$ and $v_{n+T}$, where $T = 87$. A dashed line indicates a reference line with slope equal to 1. (b) The probability density $P_{M'(x)}$ for $M'(x)$.}
\end{figure}

Similar to the analysis of the experiment, we construct a time-delayed return map for $v(t)$ using Eq. (\ref{eq:Map}). Figure \ref{fig:Exp_Maps_SIM}a shows that the dynamics of this return map form an attractor with two linear clusters that show qualitative agreement with the experiment, where the clusters produced by the model are less scattered.

These quantitative differences occur because our model does not include noise and assumes the gain to be always ideally and linearly related to $g(t)$. Nevertheless, in both the experiment and model, the return map transitions between the $g_{\text{L}}$ and $g_{\text{H}}$ clusters as $s(t)$ switches states. 

We also examine the simulated probability density $P_{M'(x)}$, given in Fig. \ref{fig:Exp_Maps_SIM}b. It shows two primary peaks located at $M'(x) \sim g_{\text{L,H}}$. Thus, the simulated dynamics also have high probabilities for the gain to satisfy $g_n = g_{\text{L,H}}$ and little probability for other values. Different from the experiment, these two peaks have smaller widths due to the assumptions of the model.  

Lastly, we use the simulated $P_{M'(x)}$ and Eq. (\ref{eq:Lyapunov}) to calculate the map's Lyapunov exponent. We calculate $h = 5.6 \times 10^{-3}$, and the scaled $h \rightarrow h/T = 6.4 \times 10^{-5}$. We note that these values are close to the Lyapunov exponents determined from the experimental data and confirm the presence of chaos in the model. 

Taken together, the time series analyses and return map demonstrate that our simple model captures the fundamental aspects of the experimental system.

\section{Conclusions and Future Directions}
In conclusion, we demonstrate a UHF electronic device displaying PWL chaos using feedback and control loops that overcomes many of the hindrances of slower LRC oscillators. We exploit a simple VGA with time delayed feedback to generate growing and decaying multi-mode oscillations centered at frequency ${\sim}2.1$ GHz. Using matched time delays for the feedback and control loops, our circuit performs fast control and avoids nonlinear saturation. As for the dynamics, this novel circuit generates simultaneously a chaotic signal with a corresponding switching state that partitions the return map's attractor, thus demonstrating a UHF foundation for the concepts of PWL chaos with a real-time symbolic dynamics.\cite{Corron2010,Corron2012}

We presented our experimental findings alongside a numerical model. To capture the key aspects from the experimental dynamics, we measure the circuit's time-delays and frequency dependent band-pass characteristics. We simulate the dynamics using delay differential equations involving a logarithmic function, a low-pass filter, and continuous switching functions to approximate the piecewise-constant parameters. Our model demonstrates a fundamental understanding of the system dynamics and gives insight into the non-ideal effects arising in the experiment. 

In the future, we are interested in applying this fully-electronic design in a radar system. Our chaotic system automatically produces an upconverted, broadband analog waveform, and information about this signal can be stored with one-bit digital sampling of $s(t)$. Preliminary analysis shows that the waveforms $v(t)$ and $s(t)$ show potential as radar signal sources when compared to standard techniques.\cite{Hall2012} Futhermore, based on the PWL nature of this system, we are currently investigating a matched \cite{Corron2010} (or pseudo-matched \cite{Cohen2012}) filter for chaos that can recover $s(t)$ from $v(t)$ in the presence of large noise. The simplicity of our design combined with these benefits represent a significant step towards inexpensive and robust chaos-based radars.  

Finally, we present only the simples case for our system where $\tau_{f}$ and $\tau_{c}$ are approximately matched. Preliminary experimental and numerical work shows that additional mismatch between these two delays can give rise to new chaotic dynamics with differently structured PSD's. Thus, the delay mismatch represents an unexplored degree-of-freedom for controlling the dynamics and exploring the stability of this system.\cite{Tsubone1998,Vu2010} Potential applications for this additional control include realizing a chaos communications system\cite{Hayes1993, Bolt2003} and creating orthogonal communications channels.\cite{Rontani2011}  

\section{ACKNOWLEDGEMENTS}
We gratefully acknowledge the financial support of Propagation Research Associates (PRA) Grant No. W31P4Q-11-C-0279.

\appendix

\section{Control Loop Characterization and Latency Time}

In this section, we characterize the three main components of the control loop: the LGA, the TTL gate, and the voltage control port $v_{\text{ctl}}$ of the VGA, in order to estimate model parameters and approximate each component's control latency time (CLT). 

\begin{figure}[t!]
\begin{center}
 \resizebox{7.5cm}{!}{\includegraphics{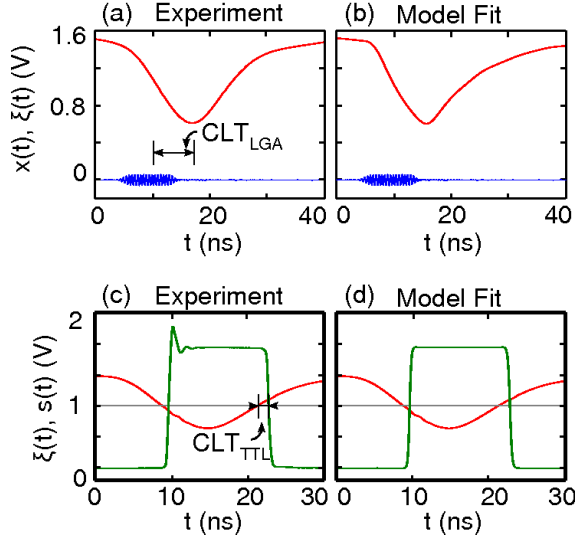}}
\end{center}
\caption{\label{fig:CLT} Control loop fits and latency times. (a) Output $\xi(t)$ (top) of the LGA due to $x(t)$ (bottom). The $\text{CLT}_{\text{LGA}}$ is shown as a delay time between the two waveforms. (b) Model fit $\xi(t)$ (top) due to input experimental $x(t)$ (bottom). (c) Example output $s(t)$ (green) of the TTL gate due to an input $\xi(t)$ from the LGA. The TTL gate threshold is given by the horizontal bar. A typical $\text{CLT}_{\text{TTL}}$ is shown for the switching in $s(t)$ from high to low. (d) Model fit $s(t)$ due to input experimental waveform $\xi(t)$.}
\end{figure}

In Fig. \ref{fig:CLT}a, we plot a typical output waveform from the LGA. Using a signal generator, we drive the LGA with a 2 GHz sinusoidal signal $x(t)$ that is pulsed for 10 ns. When zero power is present in $x(t)$, the LGA outputs a positive voltage $\xi(t) \sim 1.5$ V. For oscillations in $x(t)$, $\xi(t)$ decreases to a value that is proportional to the amplitude of $x(t)$. Hence, the output of the LGA inversely follows the envelope of input oscillations. These time series are measured using two high-impedance probes with identical propagation times after calibration. Thus, the time skew between $x(t)$ and $\xi(t)$ accurately represents the causal behavior of the LGA and shows that the $\text{CLT}_{\text{LGA}}$ is approximately 8 ns. 

Using Eqs. (\ref{eq:F}), (\ref{eq:Log_Amp_Eq}), we drive our model for the LGA using $x(t)$ and fit the output $\xi(t)$ with the model parameters $C_{i}$ (for $i  = 1,2,3$) and $f_{\text{L}}$ using a reduced $\chi^2$ algorithm.  We note that during the fitting, the CLT of our LGA model is not 8 ns due to the choice of a low-order filter. Thus, extra delay was added into the model LGA's CLT. This extra delay will be discussed at the end of this section. The resulting fit is shown in Fig. \ref{fig:CLT}b. 

We next characterize the TTL gate using the output from the LGA when it is driven by $x(t)$. Using the same high-impedance probes, we measure and plot the TTL output $s(t)$ alongside its input $\xi(t)$ in Fig. \ref{fig:CLT}c. In the figure, $s(t)$ switches high (low) for $\xi(t)$ below (above) the TTL threshold of ${\sim}1\un{V}$. We mark $\text{CLT}_{\text{TTL}} = $ 1.1 ns, the average delayed response of the gate switchings (rise and fall) in response to a threshold crossing.

We use Eq. (\ref{eq:S}) to fit the TTL gate parameters $A$, $m$, and $I$. The simple model does not produce the correct CLT for the TTL gate, as an additional 1.1 ns time delay is necessary to achieve a good fit (see discussion at the end of this section). The fit is shown in Fig. \ref{fig:CLT}d. 

Finally, we analyze the processing time of the voltage control port of the VGA in reponse to a signal from the TTL gate. To do so, we drive the VGA with a 2 GHz low-amplitude sinusoidal signal and observe the output as $s(t)$ switches to measure $\text{CLT}_{v_{\text{ctl}}} \sim 4\un{ns}$. Since the model for the gain of the VGA does not include a processing time, additional delay is necessary in the model.

Therefore, in order to match and model the full control loop delay time $\tau_{\text{c}}$, we use the relation $\tau_{\text{c}} = \text{CLT} + \tau_{\text{coax}}$, where $\text{CLT} = \text{CLT}_{\text{LGA}}+\text{CLT}_{\text{TTL}}+\text{CLT}_{v_{\text{ctl}}}$. We model the coaxial cable delay as a time shift that compensates for any CLT that is not present in our models for the LGA, TTL gate, or voltage control port. As a result, we tune the value of $\tau_c$ in the model to match that of the experiment, which is approximately 40 ns.

\section{Multi-mode feedback transfer function and fitting}

We determine the value of $f^{(+,-)}_{\text{L,H}}$ (used in Eq. (\ref{eq:piecewise_parameters_1})) using a fit to the experimental spectral responses shown in Figs. \ref{fig:Experimental_Setup}b-c. To do so, we derive the transfer function for the setup shown in Fig. \ref{fig:Experimental_Setup}a using the differential equation
\begin{equation}
\frac{\dot{v}(t)}{\Delta_{\text{L,H}}} + v(t)+\frac{\omega^{(o)}_{\text{L,H}}\int \! v(t') \, \mathrm{d} t'}{\Delta_{\text{L,H}}}=a_{\text{L,H}}(v(t-\tau_{\text{f}})+v_{\text{in}}),
\label{eq:BP_feedback_2}
\end{equation}
where $a_{\text{L,H}} = g_{a} g_{\text{L,H}}$ is the attenuated gain (to keep the system from saturating) and $\Delta_{\text{L,H}}$, $\omega^{(o)}_{\text{L,H}}$ are the bandpass filter parameters for low (L) and high (H) gain states. We note that this equation is similar to that Eq. (\ref{eq:BP_feedback}) with an additional input driving term $v_{\text{in}}$ and an attenuation $g_{\text{a}}$ in the feedback. We then Fourier transform Eq. (\ref{eq:BP_feedback_2}) and solve for $H_{\text{L,H}}(f) = \tilde{v}/\tilde{v}_{\text{in}}$, the system's transfer function, where $\tilde{v}$ and $\tilde{v}_{\text{in}}$ are the Fourier transforms of $v(t)$ and $v_{\text{in}}(t)$, respectively. The functional form of $H_{\text{L,H}}(f)$ is
\begin{equation}
H_{\text{L,H}}(f) = \frac{a_{\text{L,H}}\Delta_{\text{L,H}}}{2 \pi i f + \Delta_{\text{L,H}} + \frac{(2 \pi i f^{(o)}_{\text{L,H}})^2}{2 \pi i f} - a_{\text{L,H}}\Delta_{\text{L,H}} e^{2 \pi i f \tau_{\text{f}}}},
\label{eq:BP_Filter}
\end{equation}
where $\Delta_{\text{L,H}}$ = $2 \pi(f^{(+)}_{\text{L,H}} - f^{(-)}_{\text{L,H}})$,  and $f^{(o)}_{\text{L,H}} = \sqrt{f^{(+)}_{\text{L,H}}  f^{(-)}_{\text{L,H}}}$.

\begin{figure}[t!]
\begin{center}
 \resizebox{8cm}{!}{\includegraphics{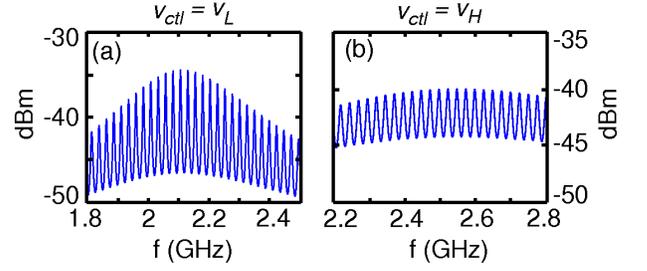}}
\end{center}
\caption{\label{fig:Spectral_Responses} Using Eq. (\ref{eq:BP_Filter}), we fit the experimental spectra measured in Fig. \ref{fig:Experimental_Setup} for (a) $v_{\text{ctl}}=v_{\text{L}}$ and (b) $v_{\text{ctl}}=v_{\text{H}}$.} 
\end{figure}

Using the magnitude of $H_{\text{L,H}}(f)$, we fit the experimental data and obtain the fits shown in Figs. \ref{fig:Spectral_Responses}a-b. The fitting parameters are $f^{(+)}_{\text{H}} = 2.41 \pm 0.01 \un{GHz}$, $f^{(-)}_{\text{H}} = 1.85 \pm 0.01 \un{GHz}$, $f^{(+)}_{\text{L}} = 3.29 \pm 0.1 \un{GHz}$, and $f^{(-)}_{\text{L}} = 1.97 \pm 0.04 \un{GHz}$. We note that the uncertainties are higher for the frequency parameters $f^{(+,-)}_{\text{L}}$ due to approximation of the resonances in Fig. \ref{fig:Experimental_Setup}c using Eq. (\ref{eq:BP_Filter}). The experimental system is not a true band-pass filter for $v_{\text{ctl}}=v_{\text{H}}$, and our approximation leads to discrepancies between Fig. \ref{fig:Experimental_Setup}c and Fig. \ref{fig:Spectral_Responses}b, but this approximation is necessary for our simple model. The fitted gain parameters are $a_{\text{H}} = 0.6 \pm 0.01$ and $a_{\text{L}}= 0.2 \pm 0.01$. Also, in order to facilitate fitting, $\tau_{f}$ is a free fitting parameter, where its fitted values are $\tau_{f} = 41 \pm 1 \un{ns}$. 

\section{Transient dynamics}

\begin{figure}[h!]
\begin{center}
 \resizebox{5.5cm}{!}{\includegraphics{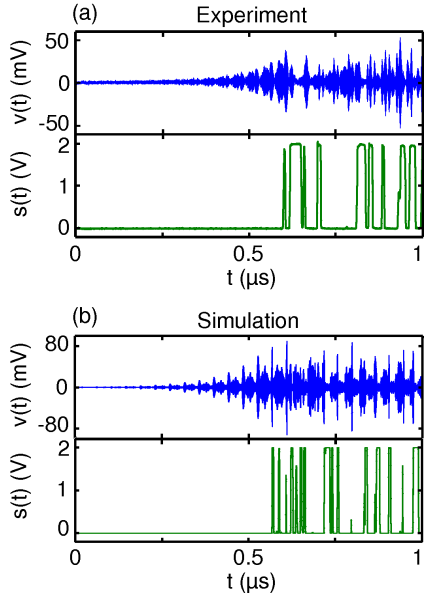}}
\end{center}
\caption{\label{fig:Transients} Transients in $v(t)$ and $s(t)$ for the (a) experiment and (b) simulation.}
\end{figure}

We plot typical transient dynamics for $v(t)$ and $s(t)$ in Fig. \ref{fig:Transients}. Electrical and additive noise seed the experiment (Fig. \ref{fig:Transients}a) and simulation (Fig. \ref{fig:Transients}b), respectively, which show multi-mode growth in $v(t)$ and that $s(t)$ begins switching at $0.5$ $\mu$s $<$ t $<0.75$ $\mu$s.

\section{Underlying Grammar}
\begin{figure}[h!]
\begin{center}
 \resizebox{8.5cm}{!}{\includegraphics{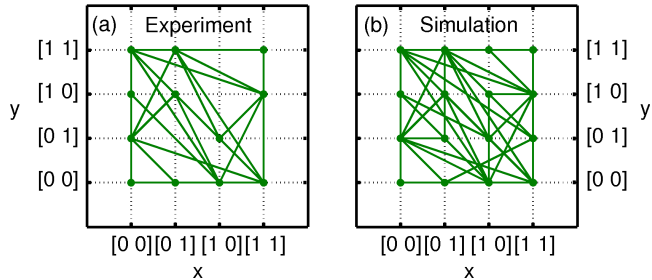}}
\end{center}
\caption{\label{fig:word_maps} Four-bit digital word maps for $10^4$ words in the (a) experimental and (b) simulated $s_{n}$.}
\end{figure}

To demonstrate the existence of a potential grammar in the experimental and simulated switching states, we examine four-bit digital words that appear sequentially in the time series for $s(t)$. To do so, we sample $s(t)$ at a fixed clock frequency of 1 GHz and convert the digital-like signal into a list of digital ones and zeros, labaled $s_{n}$: for $s(t_n) > 1, s_{n} = 1$ and for $s(t_n) < 1, s_{n} = 0$, where $t_n$ are the clock sampling times. Next, we construct the pairings [0,0], [0,1], [1,0], and [1,1] by pairing [$s_n,s_{n+1}$], [$s_{n+2},s_{n+3}$], ... and then follow the evolution of the pairs. We depict the sequences of four-bit words in Fig. \ref{fig:word_maps}, where a point is drawn between two pairings on the ($x,y$) axes if the four-bit word ([$s_n,s_{n+1}$], [$s_{n+2},s_{n+3}$]) exists in the alphabet. Lines connect sequential four-bit words.

In the experimental analysis shown in Fig. \ref{fig:word_maps}a, there are both missing words and missing links, showing a restricted alphabet and grammar. In the analysis of the simulation (Fig. \ref{fig:word_maps}b), all four-bit words are present in the alphabet but with a restricted grammar. The grammar of the model is sensitive to parameter changes, and we conjecture that mismatches with the experimental parameters, such as the order of the low-pass filter in the LGA, lead to the differences between Fig. \ref{fig:word_maps}a and Fig. \ref{fig:word_maps}b. We also note that overlapping lines are neglected and would show even more missing links in a higher-dimensional pictoral representation.

\bibliographystyle{apsrev4-1.bst}

%

\end{document}